\documentclass[twocolumn]{aastex63}
\received{2021 Aug 28}
\revised{2021 Oct 11, 21}
\accepted{2021 Oct 22}
\submitjournal{PSJ}

\shorttitle{Near-UV Reddening Observed in the Spectrum of 2012~DR$_{30}$}
\shortauthors{Seccull, Fraser, \& Puzia}

\begin{document}

\title{Near-UV Reddening Observed in the Reflectance Spectrum of High Inclination Centaur 2012~DR$_{30}$}

\correspondingauthor{Tom Seccull}
\email{tom.seccull@noirlab.edu}

\author[0000-0001-5605-1702]{Tom Seccull}
\affiliation{Gemini Observatory/NSF's NOIRLab, 670 N. A'ohoku Place, Hilo, HI 96720, USA}
\affiliation{Astrophysics Research Centre, Queen's University Belfast, University Road, Belfast, BT7 1NN, UK}

\author[0000-0001-6680-6558]{Wesley C. Fraser}
\affiliation{Herzberg Institute of Astrophysics, 5071 West Saanich Road, Victoria, BC V9E 2E7, Canada}

\author[0000-0003-0350-7061]{Thomas H. Puzia}
\affiliation{Institute of Astrophysics, Pontificia Universidad Cat\'olica de Chile, Av. Vicu\~na MacKenna 4860, 7820436, Santiago, Chile}



\begin{abstract}
Centaurs with high orbital inclinations and perihelia ($i>60^{\circ}$; $q\gtrsim15~$au) are a small group of poorly understood minor planets that are predicted to enter the giant planet region of the Solar System from the inner Oort Cloud. As such they are one of the few samples of relatively unaltered Oort Cloud material that can currently be directly observed. Here we present two new reflectance spectra of one of the largest of these objects, 2012~DR$_{30}$, in order to constrain its color and surface composition. Contrary to reports that 2012~DR$_{30}$ has variable optical color, we find that consistent measurements of its spectral gradient from most new and published datasets at $0.55-0.8~\mu$m agree with a spectral gradient of $S'\simeq10\pm1~\%/0.1~\mu$m within their uncertainties. The spectral variability of 2012~DR$_{30}$ at Near-UV/blue and Near-Infrared wavelengths, however, is still relatively unconstrained; self-consistent rotationally resolved followup observations are needed to characterise any spectral variation in those regions. We tentatively confirm previous detections of water ice on the surface of 2012~DR$_{30}$, and also consistently observe a steady steepening of the gradient of its spectrum from $\lambda\sim0.6~\mu$m towards Near-UV wavelengths. Plausible surface materials responsible for the observed reddening may include ferric oxides contained within phyllosilicates, and aromatic refractory organics.
\end{abstract}



\section{Introduction} \label{sec:intro}
Solar System minor planets with high orbital inclinations and perihelia ($i>60^{\circ}$; $q\gtrsim15~$au) comprise a small and poorly understood population known as the High-\textit{i} High-\textit{q} (\textit{HiHq}) Centaurs \citep{2012MNRAS.420.3396B}. Their precise origin is not known, but studies of their dynamical properties generally converge towards the conclusion that they are unlikely to be sourced from the trans-Neptunian scattered disk or the Kuiper Belt; instead they probably originate from a more distant reservoir of objects such as the Oort Cloud \citep{2005MNRAS.361.1345E,2009ApJ...697L..91G,2012MNRAS.420.3396B,2013Icar..224...66V,2013AA...555A...3K,2020MNRAS.497L..46M}. 

The dynamical lifetime of a centaur correlates positively with its orbital perihelion and inclination \citep[e.g.][]{2013Icar..224...66V,2020CeMDA.132...36D}, and those of \textit{HiHq} Centaurs are stable on timescales of $\sim10^7-10^8~$years \citep[e.g.][]{2009ApJ...697L..91G,2012MNRAS.420.3396B,2013Icar..224...66V,2013AA...555A...3K,2016ApJ...827L..24C,2017MNRAS.472L...1M}. This is still short relative to the age of the Solar System, however, so the extant population of \textit{HiHq} Centaurs requires the existence of one or more dynamical mechanisms capable of resupplying it with new members \citep[e.g.][]{1997Icar..127...13L,2003AJ....126.3122T}. The combined gravitational action of the galactic tide, stellar flybys, and the ice giant planets on the orbits of Oort Cloud objects is one possible mechanism suggested to explain the existence of the observed population \citep[e.g.][]{2012MNRAS.420.3396B}, and more recently so has the gravitational influence of a distant perturbing planetary mass \citep{2015Icar..258...37G,2016ApJ...833L...3B}. The discovery that some (but not all) \textit{HiHq} Centaurs occupy a shared highly inclined orbital plane suggests that objects within that plane may be delivered to planet-crossing orbits by a shared mechanism, although none of those put forward so far were found to be capable of maintaining the observed plane for any meaningful length of time \citep{2016ApJ...827L..24C}. While the precise origins of \textit{HiHq} Centaurs and their dynamical pathways into the giant planet region remain somewhat enigmatic, characterisation of the observable population is clearly crucial to exploring the Solar System beyond the Kuiper Belt, which for the most part is presently beyond our observable reach.

To date, multiband photometry has been the main technique used for characterisation of the surfaces of \textit{HiHq} Centaurs, and half of those currently known have published optical colors. 127546~(2002~XU$_{93}$), 528219~(2008~KV$_{42}$), 2010~WG$_{9}$, and 2012~DR$_{30}$ have all been reported to have grey to moderately red optical colors similar to those of cometary nuclei, the Damocloids, the Trojan asteroids of Jupiter and Neptune, the less-red Trans-Neptunian Objects (TNOs), and the primitive X- and D-type asteroids \citep{2010AJ....139.1394S,2012AA...546A.115H,2013AA...555A...3K,2013AJ....146...17R,2015AJ....150..201J,2018AJ....155..170S,2021AA...647A..71H}. The similarity between the colors of these minor planet populations and those of \textit{HiHq} Centaurs suggests that the \textit{HiHq} Centaurs are likely to have dark and carbonaceous surfaces that may have experienced significant irradiation and thermal evolution \citep[e.g.][]{2010AJ....139.1394S}.

Of the known \textit{HiHq} Centaurs, 2012~DR$_{30}$ is the most thoroughly characterised, and has the most extreme orbit ($a=1334~$au; $e=0.98909$, $i=78^{\circ}$; $q=14.56~$au; $Q=2654~$au). Both \citet{2013AA...555A...3K} and \citet{2018AJ....155..170S} noted that due to its high aphelion and the orientation of its orbit away from the heliotail toward an ecliptic longitude of $\sim340^{\circ}$, 2012~DR$_{30}$ spends most of its time beyond the heliosphere in local interstellar space where it experiences strong cosmic particle irradiation. The orbit of 2012~DR$_{30}$ is not within the common plane identified by \citet{2016ApJ...827L..24C}, so its orbit may have become planet-crossing by a different mechanism to those that are.

With a radiometrically determined effective diameter of $\sim185~$km, 2012~DR$_{30}$ is the largest of the \textit{HiHq} Centaurs that have been observed at thermal wavelengths \citep[for 2002~XU$_{93}$ and 2010~WG$_{9}$ see][]{2012A&A...541A..92S,2013ApJ...773...22B}, and indeed is large in comparison to most centaurs, which typically have diameters below 120~km \citep{2014A&A...564A..92D}. The radiometrically determined geometric albedo of 2012~DR$_{30}$ is $p_V\sim8\%$ \citep{2013AA...555A...3K}, which is consistent with those of centaurs, but lies at the higher end of the centaur albedo distribution \citep{2014A&A...564A..92D}. \citet{2013AA...555A...3K} noted that the beaming parameter of 2012~DR$_{30}$, at $\eta\simeq0.81$, is closer to those typical of main belt asteroids \citep[e.g. see][]{2011ApJ...741...68M,2011ApJ...743..156M,2012ApJ...744..197G,2018A&A...612A..85A} than those of TNOs \citep[e.g.][]{2012A&A...541A..92S,2012A&A...541A..93M,2012A&A...541A..94V,2013A&A...557A..60L}, but also that this value still falls within the broad range of $\eta$~values occupied by the centaurs \citep{2014A&A...564A..92D}. 

The optical and Near-Infrared (NIR) reflectance spectra of 2012~DR$_{30}$ reported by \citet{2018AJ....155..170S} were observed to be featureless, red, and broadly linear across the range $0.4-1.7~\mu$m. The only absorption band detected by \citet{2018AJ....155..170S} was a $2.0~\mu$m band characteristic of the presence of surface water ice; a weaker water ice absorption band at $\lambda\sim1.55~\mu$m that often accompanies the band at $2.0~\mu$m was not detected. A potential silicate absorption band predicted to be present at $\lambda\sim0.9~\mu$m by \citet{2013AA...555A...3K} was also not detected. Spectral modeling by \citet{2018AJ....155..170S} suggested that the reflectance properties of 2012~DR$_{30}$ could be approximated by that of a mixture of water ice (both with and without organic inclusions), Titan and Triton tholins, and pyroxene; the addition of amorphous carbon to this mixture was found to decrease consistency between the model spectrum and the observed one. 

As a collection, the optical colors reported for 2012~DR$_{30}$ so far suggest that its color varies significantly between observational epochs \citep{2013AA...555A...3K,2018AJ....155..170S,2021AA...647A..71H}.

Our team spectroscopically observed 2012~DR$_{30}$ in 2015 and 2017. Here we present two new reflectance spectra of 2012~DR$_{30}$ resulting from those observations that we use to constrain its color and surface composition.

\section{Observations}\label{sec:obs}

\begin{table*}
\footnotesize
\centering
\caption{Observation Log}
\begin{tabular}{ p{1.9cm}p{4.1cm}p{0.7cm}p{0.7cm}p{0.7cm}p{0.4cm}p{1.7cm}p{1.4cm}p{1.4cm}p{1.4cm}}
\hline\hline
Target & Observation Date $\vert$ UT Time & \multicolumn{3}{c}{T\textsubscript{exp}, s} & N\textsubscript{exp} & Airmass & \multicolumn{3}{c}{IQ, \arcsec} \\[2pt]
\hline
\multicolumn{2}{l}{VLT/X-Shooter} & UVB & VIS & NIR & ~ & ~ & UVB & VIS & NIR\\
\hline
HD 76440 & 2015 Jun 10 $\vert$ 23:32:51--23:39:12 & 13.0 & 13.0 & 13.0 & 3 & 1.421--1.426 & 1.41--1.68 & 1.06--1.21 & 0.93--1.10\\[1pt]
2012~DR$_{30}$ & 2015 Jun 10 $\vert$ 23:53:55--01:32:16 & 500.0 & 466.0 & 612.0\textsuperscript{\textit{a}} & 8 & 1.246--1.654 & 1.27--1.44 & 1.04--1.19 & 0.80--0.91\\[1pt]
\textbf{HD 97356} & 2015 Jun 11 $\vert$ 02:12:38--02:18:55 & 10.0 & 10.0 & 10.0 & 3 & 1.817--1.832 & 1.43--1.68 & 1.07--1.26 & 0.89--1.06\\[1pt]

\hline
\multicolumn{2}{l}{VLT/FORS2} & ~ & ~ & ~ & ~ & ~ & ~ & ~ & ~\\ 
\hline
HD 69216 & 2017 Apr 21 $\vert$ 23:12:55--23:21:31 & ~ & 3.0 & ~ & 2 & 1.169 & ~ & 0.65--0.69 & ~\\[1pt]
HD 87027 & 2017 Apr 21 $\vert$ 23:22:03--23:28:28 & ~ & 3.0 & ~ & 2 & 1.177 & ~ & 0.75--0.88 & ~\\[1pt]
2012~DR$_{30}$ & 2017 Apr 21 $\vert$ 23:28:47--01:02:59 & ~ & 500.0 & ~ & 8 & 1.175--1.197 & ~ & 0.68--1.15 & ~\\[1pt]
HD 100076 & 2017 Apr 22 $\vert$ 01:09:47--01:17:03 & ~ & 3.0 & ~ & 2 & 1.149 & ~ & 0.71--0.78 & ~\\[1pt]
HD 101082 & 2017 Apr 22 $\vert$ 01:17:08--01:22:54 & ~ & 2.0 & ~ & 2 & 1.164 & ~ & 0.68--0.78 & ~\\[1pt]
\textbf{BD-00 2514} & 2017 Apr 22 $\vert$ 01:30:52--01:41:46 & ~ & 12.0 & ~ & 2 & 1.127 & ~ & 0.65--0.69 & ~\\[1pt]
\hline
\end{tabular}\\[6pt]
\small{\textbf{Note.} For each target we present the UT observation date and time, the integration time per exposure (T\textsubscript{exp}), the number of exposures per wavelength range observed (N\textsubscript{exp}), and the airmass at which they were observed. The range of estimated image quality (IQ) values presented for each target is the range of Full Widths at Half Maximum measured from Moffat profiles \citep{1969A&A.....3..455M} fitted to the set of spatial profiles produced by median collapsing each of a target's reduced 2D spectroscopic exposures along the dispersion axis. Spectra of the stars presented in bold were the ones ultimately used to calibrate the reflectance spectra of 2012~DR$_{30}$.\\
\textsuperscript{\textit{a}} Total exposure time reached by coadding two 306~s integrations.}
\label{tab:log}
\end{table*}

\begin{table*}
\centering
\caption{Observation Geometry}
\label{tab:obsdets}
\begin{tabular}{p{2.3cm}p{1.4cm}p{1.5cm}p{1.cm}p{0.8cm}p{0.8cm}p{1.8cm}}
\hline\hline
Instrument & R.A., hr & Decl, $^{\circ}$ & $r_{H}$, au & $\Delta$, au & $\alpha$, $^{\circ}$ & $m_V$, mag \\[2pt]
\hline
VLT/X-Shooter & 09~05~04.91 & -41~24~25.5 & 16.083 & 16.042 & 3.6186 & $19.59\pm0.36$\\[1pt]
VLT/FORS2 & 09~24~01.75 & -56~14~37.2 & 17.540 & 17.141 & 3.0494 & $19.85\pm0.36$\\[1pt]
\hline
\end{tabular}\\[6pt]
\small{\textbf{Note.} Values presented here correspond to the temporal median of our observations of 2012~DR$_{30}$. Alongside the instrument used during each observation epoch we present the coordinates of 2012~DR$_{30}$, heliocentric distance ($r_H$), geocentric distance ($\Delta$), solar phase angle ($\alpha$), and apparent magnitude ($m_V$). Apparent magnitudes were calculated using $m_V = H_V + 5log_{10}(r_H\Delta)+\beta\alpha$, where $H_V=7.04\pm0.35~$mag and $\beta=0.137\pm0.09~$mag$/^{\circ}$~\citep{2013AA...555A...3K}.}
\label{tab:geom}
\end{table*}

We observed 2012~DR$_{30}$ during two separate visitor mode runs at the European Southern Observatory's (ESO's) Very Large Telescope (VLT). Observing logs and the observation geometry of 2012~DR$_{30}$ at each epoch are respectively presented in Tables \ref{tab:log} and \ref{tab:geom}. Both observing runs were conducted under clear (sky $90\%$ cloudless at elevation $> 30^{\circ}$, transmission variation $< 10\%$) or photometric (sky cloudless, transmission variation $< 2\%$) conditions as defined by ESO; as a result we expect the continuum colors of our spectra to be minimally affected by clouds, and therefore accurate. As in previously reported observational epochs \citep{2013AA...555A...3K,2018AJ....155..170S,2021AA...647A..71H}, 2012~DR$_{30}$ was found to be inactive both times we observed it.

\subsection{X-Shooter}
During the first run, on 2015 Jun 10-11 UT, we observed 2012~DR$30$ with the X-Shooter spectrograph, mounted on the 8.2m UT2(Kueyen) unit telescope. X-Shooter is a medium resolution echelle spectrograph with three arms covering the Near-UV/blue (UVB; $0.30-0.56~\mu$m), visual (VIS; $0.55-1.02~\mu$m), and NIR (NIR; $1.02-2.48~\mu$m) spectral ranges that can be exposed simultaneously \citep{2011A&A...536A.105V}. We did not make use of X-Shooter's $K$-band blocking filter for these observations, so the uncalibrated spectroscopic observations covered the full observable wavelength range. No detector binning was implemented, so data was obtained at the native pixel scales of each detector (UVB \& VIS, $0.16\arcsec$; NIR, $0.21\arcsec$). For the UVB, VIS, and NIR spectrographic arms we respectvely used slit widths of $1.0\arcsec$, $0.9\arcsec$, and $0.9\arcsec$, which each provided respective resolving powers ($\lambda/\Delta\lambda$) of $\sim5400$, $\sim8900$, and $\sim5600$; all three slits had a common length of $11\arcsec$.

We obtained spectra of 2012~DR$_{30}$ and two Solar twins from the catalog of \citet{2015A&A...574A.124D} adjacent in time and at a similar airmass and pointing (see Table \ref{tab:log}). We used a three-point repeating dither pattern of $0\arcsec$, $+2.5\arcsec$, and $-2.5\arcsec$ along the slit relative to the slit center when observing these targets to mitigate the effects of bad pixel artifacts and cosmic rays. To minimise slit losses resulting from atmospheric differential refraction \citep{1982PASP...94..715F}, and to account for X-Shooter's disabled atmospheric dispersion corrector (ADC), the slit was realigned to the parallactic angle after the completion of each dither cycle (every three exposures). One flux calibrator, EG~274, was observed as a part of the standard X-Shooter calibration program \citep{2010HiA....15..535V}.

\subsection{FORS2}
\begin{table}
\centering
\caption{$BVRI$ \& $gri$ Solar Colors}
\begin{tabular}{p{1.cm}p{1.75cm}p{1.75cm}p{1.75cm}}
\hline\hline
J-C & $(B-V)_{\odot}$ & $(V-R)_{\odot}$ & $(R-I)_{\odot}$ \\
\hline
~ & $0.653\pm0.003$ & $0.356\pm0.003$ & $0.345\pm0.004$\\
\hline
SDSS & $(g'-r')_{\odot}$ & $(r'-i')_{\odot}$ & ~ \\
\hline
~ & $0.44\pm0.02$ & $0.11\pm0.02$ & ~\\
\hline
\end{tabular}\\[6pt]
\small{\textbf{Note.} Table presenting Solar colors reported by \textcite{2012ApJ...752....5R} and SDSS for the Johnson-Cousins (J-C) and SDSS filter systems respectively.}
\label{tab:solcols}
\end{table}

We revisited 2012~DR$_{30}$ on 2017 Apr 21-22 UT with the FOcal Reducer/low dispersion Spectrograph 2 \citep[FORS2;][]{1998Msngr..94....1A}, which is mounted on the 8.2~m UT1(Antu) unit telescope. FORS2 was configured in longslit spectroscopy mode, with the blue-sensitive E2V CCD detector, the standard resolution collimator, and a slit width of $0.7\arcsec$. We used the \verb:GRIS_600B+22: grism, which provided a wavelength coverage of $0.330-0.621~\mu$m and a resolving power of $\lambda/\Delta\lambda\sim1100$ at $0.463~\mu$m when used with our chosen slit.

We observed spectra of 2012~DR$_{30}$ and five solar calibrator sources close in time and at similar airmass and pointing (see Table \ref{tab:log}). The high throughput and sensitivity of VLT/FORS2 made selection of solar calibrator stars challenging, as most well characterised solar twins and analogs were capable of saturating even the shortest spectroscopic exposures due to their brightness. Prior to our observing run three of our stars were selected because they were sufficiently faint, near to 2012~DR$_{30}$ at the time of observation, and listed with G2V spectral type and $B-V$ photometric colors within 0.02 magnitudes of Solar $B-V$ in the SIMBAD catalog \citep[][ see Table \ref{tab:solcols} for Solar colors]{2000A&AS..143....9W}. A fourth star, HD~101082, had G0V spectral type and $B-V=0.57\pm0.02$, but was chosen as a calibrator-of-last-resort in the event that others were unusable. A fifth faint calibrator star was later selected while we were at the observatory from a list of stars with precisely measured near-Solar $UBVRI$ colors compiled by Brian A. Skiff from catalogs of standard stars. In the end it was the spectrum of this fifth star, BD-00~2514 \citep[a.k.a. SA~103-204; for colors see][]{1986ApJS...60..577T} that we used to calibrate our FORS2 spectrum of 2012~DR$_{30}$, because of all the calibrator stars we observed it provided the best cancellation of the Solar metal lines in the raw spectrum of 2012~DR$_{30}$.

We used a two-point repeating dither pattern for all our FORS2 observations that alternated between $+1\arcsec$ and $-1\arcsec$ along the slit from the slit center. Again, the slit was periodically realigned to the parallactic angle to minimise slit losses. All exposures were read out using the 100 kHz, $2\times2$ binned, high gain readout mode.
\newline

\section{Data Reduction}\label{sec:red}
For all our spectra from both instruments, initial data reduction steps including bias subtraction, flat field correction, wavelength calibration, and spectrum rectification (i.e. straightening of the echelle orders) were performed with the data reduction pipelines provided for X-Shooter \citep[v.2.7.1;][]{2010SPIE.7737E..28M} and FORS2 (v.5.2.3\footnote{\url{ftp://ftp.eso.org/pub/dfs/pipelines/instruments/fors/fors-pipeline-manual-5.10.pdf}}) by ESO in the ESO Reflex data reduction environment \citep[v.2.8.4 \& v.2.8.5 respectively;][]{2013A&A...559A..96F}. The X-Shooter pipeline was also used for merging the echelle orders of each spectrum, and correcting the spectra for instrument response using the observed flux standard. The instrument response of FORS2 is stable over a timescale of days and was not corrected for in our data; because it would have been applied to both the raw spectrum of 2012~DR$_{30}$ and the Solar calibrator spectrum, such a correction would have been negated when the former was divided by the latter to derive the reflectance spectrum. 

Sky subtraction, cosmic-ray removal, and extraction of the spectra from both instruments were performed using a method similar to that used by \citet{2018ApJ...855L..26S}, where Moffat functions \citep{1969A&A.....3..455M} were fitted to the spatial profile of the 2D wavelength-calibrated spectrum at many locations along the dispersion axis in order to track the wavelength dependent center and width of the spectrum's spatial profile. Sky region boundaries were defined at $\pm3$ Full-Widths at Half Maximum (FWHM) from each Moffat profile center with sky outside these boundaries. 15 iterations of sigma clipping at 5$\sigma$ were separately conducted in the sky and target regions to remove cosmic rays. In each unbinned wavelength element the median background was subtracted. The Moffat fitting process was repeated on the sky-subtracted 2D spectra with extraction limits defined at $\pm2$ FWHM from each of the Moffat profile centers. During this second round, pixels that had been flagged as bad, or containing cosmic rays, had their values replaced using a process similar to that used in the ESO X-Shooter pipeline for the same purpose \citep{2010SPIE.7737E..28M}; details are presented in the X-Shooter pipeline manual\footnote{see the Standard Exraction algorithm; \url{ftp://ftp.eso.org/pub/dfs/pipelines/instruments/xshooter/xshoo-pipeline-manual-3.3.5.pdf}}. Within the defined extraction limits the flux was summed for each wavelength element to form the 1D spectrum. At this stage, the signal-to-noise ratio (S/N) of our X-Shooter spectra of 2012~DR$_{30}$ were found to be too low to be of use in analysis at $\lambda>1.8~\mu$m. This is primarily due to the relatively low sensitivity of X-Shooter in the $K-$band; our use of a long $2\times306~$s integration time per NIR X-Shooter frame (see Table \ref{tab:log}) also made accurate sky subtraction challenging in this region.

The FORS2 spectra were all then corrected for atmospheric extinction using $f_{C}(\lambda) = f(\lambda)10^{0.4ak(\lambda)}$, where $f_{C}(\lambda)$ and $f(\lambda)$ are respectively the extinction corrected and uncorrected spectra, $a$ is the median airmass at which the spectrum was observed, and $k(\lambda)$ is the FORS2 extinction coefficient (taken from the FORS2 data reduction pipeline) interpolated to the resolution of the spectra. Atmospheric extinction correction was not needed for the X-Shooter spectra as they had already been flux calibrated by the ESO pipeline.

Next, the spectra for each target observed with a given instrument were median stacked. All individual spectra of each Solar calibrator star were stacked. For 2012~DR$_{30}$, however, some spectra were omitted from stacking. One spectrum was omitted from the X-Shooter UVB stack, and two were omitted from both the X-Shooter VIS stack and NIR stack, because the first triplet observing sequence performed with X-Shooter was observed with 2012~DR$_{30}$ poorly centered in the slit; this reduced the S/N of these exposures and caused the blue end of the spectrum to fall off the slit at certain dither positions. One further frame was also left out of the X-Shooter NIR stack due to low S/N. The first three frames observing 2012~DR$_{30}$ in the FORS2 stack were also omitted, as this observation began soon after the end of evening twilight while the sky was still darkening, resulting in low S/N at short wavelengths. The remaining five frames, observed once the sky was sufficiently dark, provided better S/N into the Near-UV (NUV). The stacked calibrator star spectra were used to calibrate the signature of the Solar spectrum out of the stacked spectra of 2012~DR$_{30}$. We used the spectrum of HD~97356, rather than that of HD~76440, to calibrate our X-Shooter spectrum of 2012~DR$_{30}$, as HD~97356 was the brighter of the two calibrators observed and had a spectrum with higher S/N; as previously discussed in section \ref{sec:obs} we calibrated the FORS2 spectrum with that of BD-00~2514. The resulting reflectance spectra were binned using the bootstrapping method described by \citet{2019AJ....157...88S}. Binning factors (i.e. numbers of points per bin) of 130, 125, and 260 were respectively used when binning the spectra observed with the X-Shooter UVB, VIS, and NIR arms. A binning factor of 25 was used with the FORS2 data. The UVB, VIS, and NIR X-Shooter reflectance spectra of 2012~DR$_{30}$ were aligned to each other via a linear fit to the wavelength ranges adjacent to the gaps between each arm.

\section{Analysis}\label{sec:analysis}

\begin{figure*}
\centering
\includegraphics[scale=0.7]{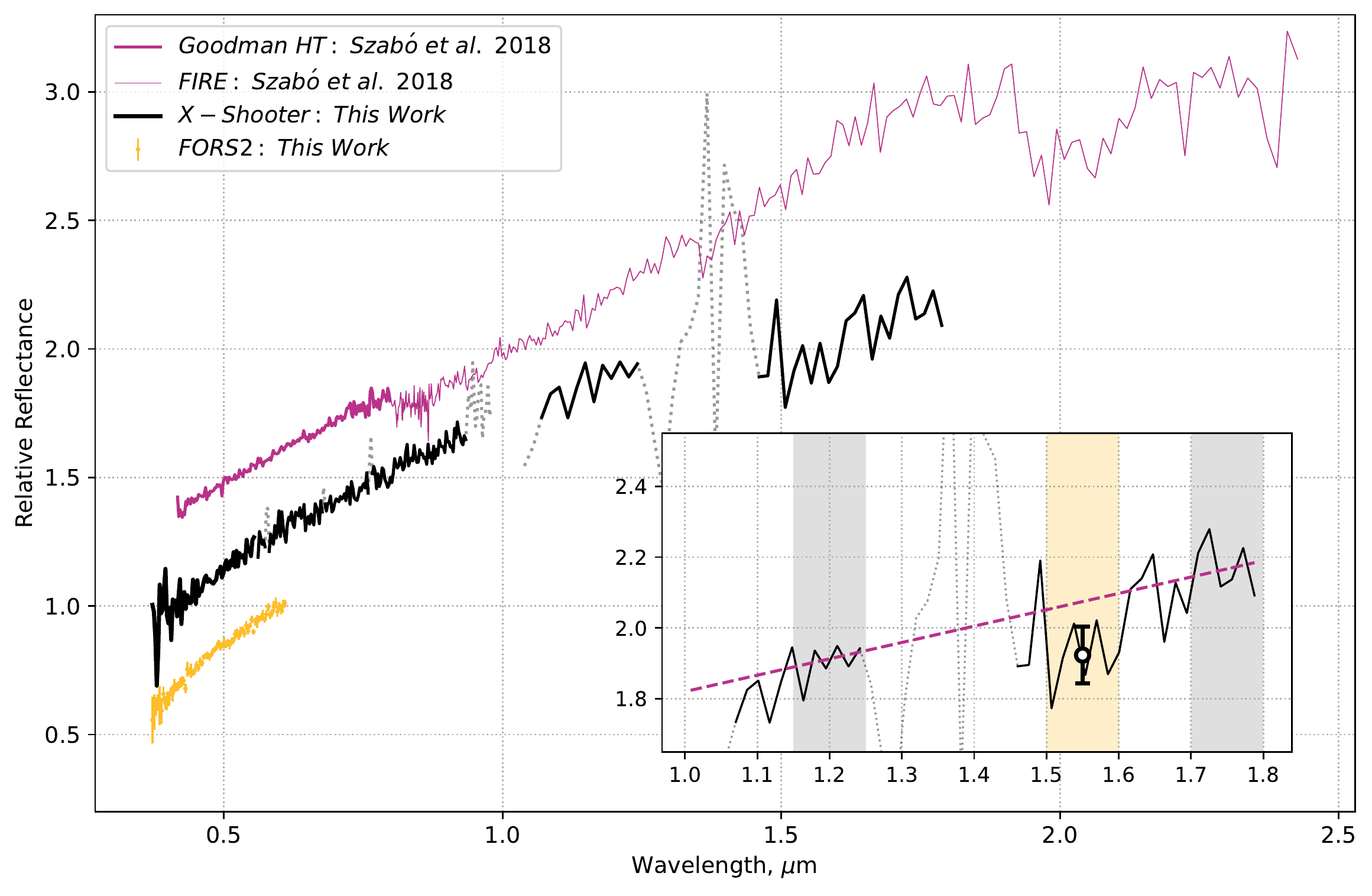}
\caption{This figure presents all currently reported reflectance spectra of 2012~DR$_{30}$. The spectra observed with the X-Shooter and FORS2 spectrographs are a product of this work, and the others were reported by \citet{2018AJ....155..170S}. All spectra have been scaled to unity at $0.6~\mu$m; the X-Shooter spectrum and the previously reported spectra are offset in increments of +0.3 for clarity. Regions of the X-Shooter spectrum contaminated with telluric residuals and instrument artifacts are plotted with dotted lines. \textbf{Inset:} This shows a zoomed in view of the NIR X-Shooter spectrum overplotted with the linear continuum fitted to the spectrum in the gray shaded wavelength regions at $1.15-1.25~\mu$m and $1.7-1.8~\mu$m. The reflectance of this modeled continuum at $1.55~\mu$m was compared to the reflectance of the spectrum averaged over the yellow shaded region at $1.5-1.6~\mu$m (represented by the point at $1.55~\mu$m) so we could estimate the depth and detection significance of the observed $1.55~\mu$m water ice band. The errorbars of the $1.55~\mu$m point representing its $1\sigma$ uncertainties. The axes of the inset are identical to those in the main plot. \label{fig:vnir}}
\end{figure*}

\begin{figure}
\centering
\includegraphics[scale=0.8]{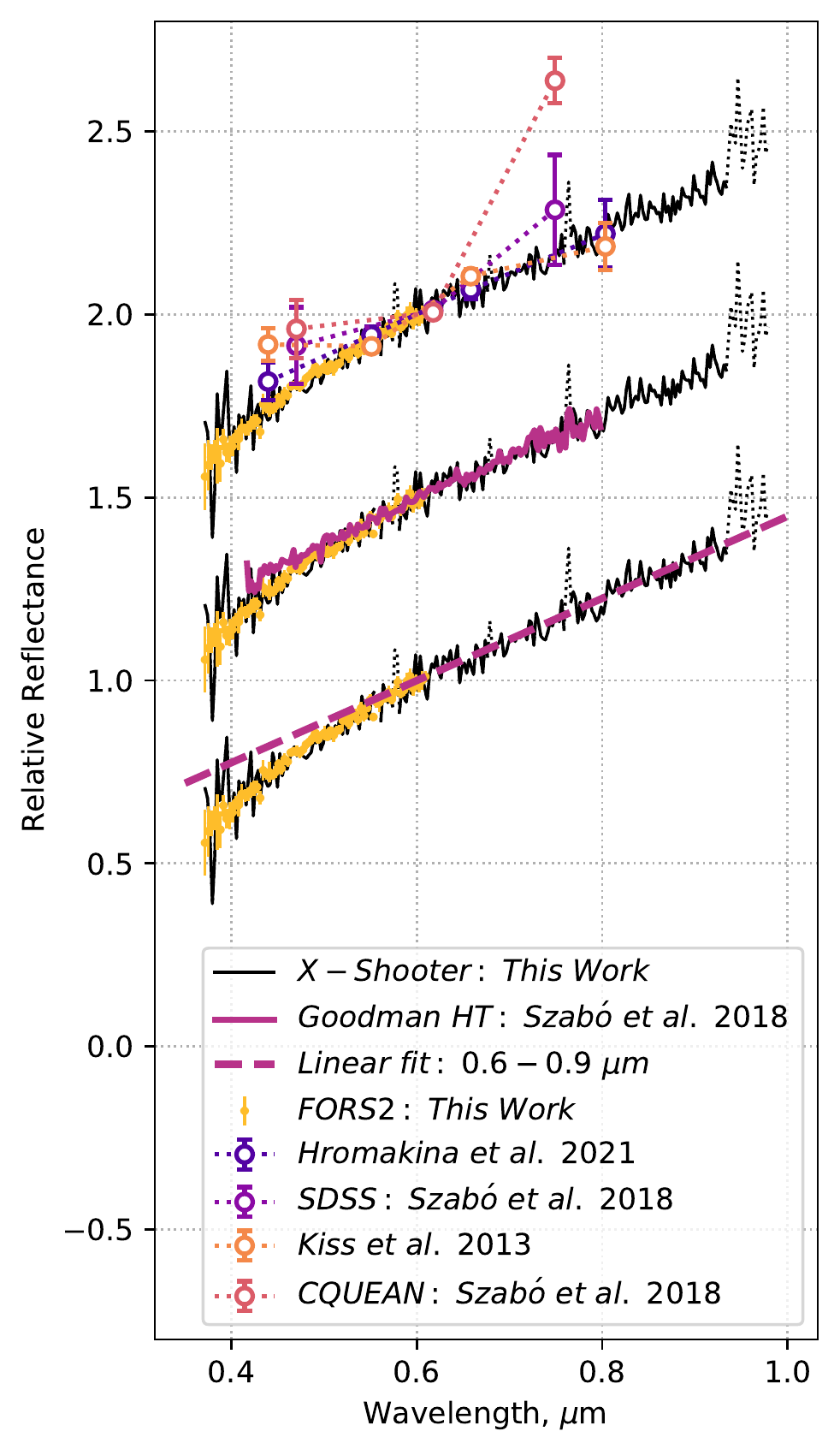}
\caption{This figure presents optical reflectance data reported for 2012~DR$_{30}$. All datasets have been scaled to unity at $0.6~\mu$m, and, where necessary, are offset for clarity in crements of +0.5.  \label{fig:vis}}
\end{figure}

\begin{figure*}
\centering
\includegraphics[scale=0.7]{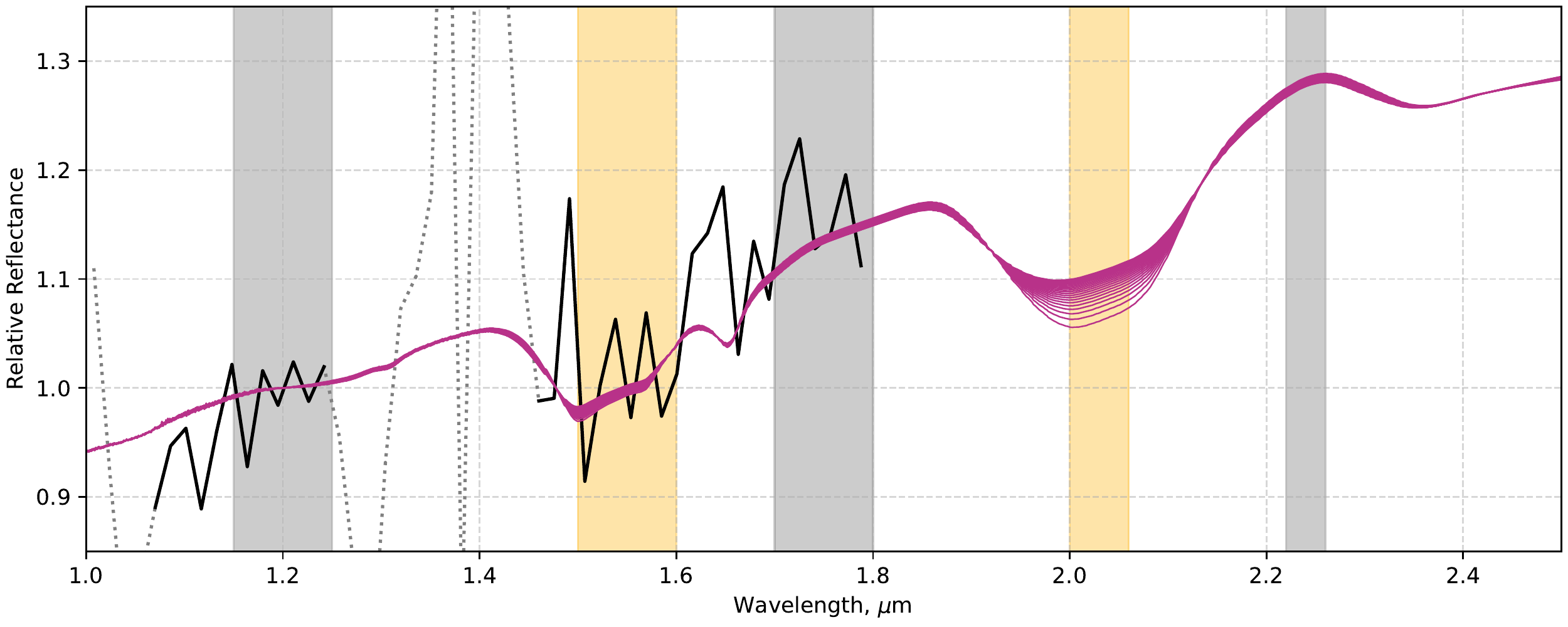}
\caption{This figure compares all model water ice reflectance spectra produced in section \ref{sec:nir} to that of 2012~DR${30}$ observed with X-Shooter, and shows the estimated range of $2.0~\mu$m band depths. The model spectra have been scaled so their $1.55~\mu$m bands have a depth of 9\% relative to the estimated continuum, and a continuum with gradient equal to that measured for the continuum fitted to $1.15-1.25~\mu$m and $1.7-1.8~\mu$m has been applied across the full NIR wavelength range. Gray shaded regions mark the parts of the model spectra used to estimate continuum reflectance, and the yellow regions show the locations used to estimate the average reflectance at the center of each band. All spectra are scaled to unity at $1.2~\mu$m. \label{fig:mod}}
\end{figure*}

We present our full optical and NIR reflectance spectra of 2012~DR$_{30}$ in Figure \ref{fig:vnir} alongside the two reflectance spectra of 2012~DR$_{30}$ presented by \citet{2018AJ....155..170S}. The optical and NIR spectra combined and published by \citet{2018AJ....155..170S} were observed at different epochs and, as mentioned in that article, likely observed different faces of 2012~DR$_{30}$; we therefore plot them in Figure \ref{fig:vnir} as they were initially published, but we differentiate the two combined spectra using lines of different thickness. 

In Figure \ref{fig:vis} we present our optical reflectance spectra of 2012~DR$_{30}$ in comparison to all published optical photometric and spectroscopic reflectance data where such data covers $0.6~\mu$m, enabling consistent scaling between datasets \citep[literature datasets include those reported by][]{2013AA...555A...3K,2018AJ....155..170S,2021AA...647A..71H}. Optical photometric colors were converted to coarse reflectance spectra using methods described by \citet{2002A&A...389..641H}, and were scaled relative to the reflectance value at $0.6~\mu$m on a line drawn between the two datapoints either side of that wavelength. $BVRI$ Solar colors from \citet{2012ApJ...752....5R} were used to convert the $BVRI$ colors of 2012~DR$_{30}$ to reflectance; likewise, $gri$ Solar colors reported on the SDSS webpages \footnote{\url{https://www.sdss.org/dr12/algorithms/ugrizvegasun/}} were used to convert the $gri$ colors of 2012~DR$_{30}$ (see Table \ref{tab:solcols}).

\subsection{Spectral Gradients}\label{sec:grads}
For the purpose of comparison we measured the gradients of our reflectance spectra of 2012~DR$_{30}$ along with those obtained or derived from the literature. The bootstrapping method described by \citet{2019AJ....157...88S} was used to measure the spectral gradient ($S'$) and estimate both the standard deviation ($\sigma$) and the standard error of the mean ($\sigma_{S'}$) of that measurement. We take $\sigma_{S'}$ to indicate the accuracy of the gradient measured from a given dataset, and use $\sigma$ to compare the consistency of gradients measured from different datasets. For our binned reflectance spectra, each datapoint was sampled by bootstrapping the sample of points in the unbinned spectrum that were combined to make it during binning. For the coarse color-derived reflectance spectra, each datapoint was (crudely) assumed to have a gaussian uncertainty distribution with a standard deviation equal to the datapoint's uncertainty; the gaussian uncertainty distribution of each datapoint was therefore randomly sampled when it was bootstrapped. It was not possible to use the bootstrapping method to measure the gradient of the reflectance spectra published by \citet{2018AJ....155..170S}, as the datafiles published with their article were done so without uncertainties. We resorted instead to measuring the gradients of these spectra with a single linear regression, which was sufficient to estimate $S'$ and $\sigma_{S'}$, but not $\sigma$. Measurements of the gradients of the reflectance spectra presented in Figures \ref{fig:vnir} and \ref{fig:vis} are presented in Table \ref{tab:grad} for multiple wavelength ranges. The gradients reported for a given wavelength range were all consistently measured relative to the reflectance at the central wavelength within that range.   

\begin{table*}
\centering
\caption{2012~DR$_{30}$ Spectral Gradients in $\%/0.1~\mu$m}
\begin{tabular}{p{0.6cm}p{0.5cm}p{0.5cm}p{1.7cm}p{1.9cm}p{1.cm}p{6.cm}}
\hline\hline
$S'$ & $\sigma$ & $\sigma_{S'}$ & $\lambda$~Range\textsuperscript{a} & Data Type\textsuperscript{b} & Bands\textsuperscript{c} & Source\textsuperscript{d} \\[2pt]
\hline
10.24 & 0.57   & 0.02 & 0.551-0.803 & Spectroscopic & $\sim$ & X-Shooter: This Work\\[1pt]
10.55 & $\sim$ & 0.06 & 0.551-0.803 & Spectroscopic & $\sim$ & Goodman-HT: \citet{2018AJ....155..170S}\\[1pt]
9.98  & 3.33   & 0.11 & 0.551-0.803 & Photometric & VRI &  \citet{2021AA...647A..71H}\\[1pt]
9.87  & 2.32   & 0.07 & 0.551-0.803 & Photometric & VRI &  \citet{2013AA...555A...3K}\\[1pt]
17.69 & 9.05   & 0.29 & 0.551-0.803 & Photometric & ri &  SDSS: \citet{2018AJ....155..170S}\\[1pt]
20.79 & 7.81   & 0.25 & 0.551-0.803 & Photometric & ri &  CQUEAN, 2013-03-23: \citet{2018AJ....155..170S}\\[1pt]
37.25 & 5.45   & 0.17 & 0.551-0.803 & Photometric & ri &  CQUEAN, 2013-04-13: \citet{2018AJ....155..170S}\\[1pt]
\hline
19.81   & 2.60   & 0.08 & 0.440-0.551 & Spectroscopic & $\sim$ &  X-Shooter: This Work\\[1pt]
19.72   & 0.89   & 0.03 & 0.440-0.551 & Spectroscopic & $\sim$ &  FORS2: This Work\\[1pt]
13.94   & $\sim$ & 0.19 & 0.440-0.551 & Spectroscopic & $\sim$ &  Goodman-HT: \citet{2018AJ....155..170S}\\[1pt]
12.99   & 5.99   & 0.19 & 0.440-0.551 & Photometric & BV &  \citet{2021AA...647A..71H}\\[1pt]
$-0.25$ & 4.62   & 0.15 & 0.440-0.551 & Photometric & BV &  \citet{2013AA...555A...3K}\\[1pt]
\hline
16.27 & 0.53   & 0.02 & 0.470-0.618 & Spectroscopic & $\sim$ &  FORS2: This work\\[1pt]
13.65 & $\sim$ & 0.08 & 0.470-0.618 & Spectroscopic & $\sim$ &  Goodman-HT: \citet{2018AJ....155..170S}\\[1pt]
7.47  & 7.77   & 0.29 & 0.470-0.618 & Photometric & gr &  SDSS: \citet{2018AJ....155..170S}\\[1pt]
3.25  & 5.45   & 0.17 & 0.470-0.618 & Photometric & gr &  CQUEAN, 2013-04-13: \citet{2018AJ....155..170S}\\[1pt]
\hline
6.12 & 1.15   & 0.04 & 1.070-1.250 & Spectroscopic & $\sim$ &  X-Shooter: This work\\[1pt]
9.28 & $\sim$ & 0.05 & 1.070-1.250 & Spectroscopic & $\sim$ &  FIRE: \citet{2018AJ....155..170S}\\[1pt]
\hline
\end{tabular}\\[6pt]
\small{\textbf{Note.} Here we present spectral gradients ($S'$), associated $1\sigma$ uncertainties ($\sigma$), and associated standard errors ($\sigma_{S'}$) measured for 2012~DR$_{30}$ across multiple wavelength ranges from our new spectra and several published spectroscopic and photometric datasets. \textbf{a:} The \textit{$\lambda$~Range} column presents the wavelength range across which each $S'$ measurement was made; all $S'$ values were measured relative to the reflectance at the center of this range. \textbf{b:} The \textit{Data Type} column shows whether each measurement was derived from spectroscopic or photometric data. \textbf{c:} The \textit{Bands} column shows, if the data type is photometric, which bands within a photometric dataset were used when deriving a given $S'$ measurement. \textbf{d:} The \textit{Source} column notes the articles in which each dataset has been published and, where an article reports multiple datasets, the instrument used to obtain it.}\\
\label{tab:grad}
\end{table*}

\subsection{NIR Absorption Bands}\label{sec:nir}

One absorption band was identified in the X-Shooter spectrum of 2012~DR$_{30}$, centered at $\lambda\sim1.55~\mu$m (see the inset in Figure \ref{fig:vnir}). Based on its shape and location, it is most likely to be attributable to the presence of water ice on the surface of 2012~DR$_{30}$. The depth of this band, $D_W$, was estimated by measuring the average reflectance of 2012~DR$_{30}$ within the band at $1.5-1.6~\mu$m, $R_B$, relative to an estimate of the continuum reflectance at $1.55~\mu$m, $R_C$. A model linear continuum slope was fitted to regions of reliable continuum on each side of the band at $1.15-1.25~\mu$m and $1.7-1.8~\mu$m using the bootstrapping method previously used to obtain spectral gradients in section \ref{sec:grads}. Use of this method provided both an average value for $R_C$ and an uncertainty. The uncertainty of $R_B$ was estimated to be the standard deviation of the reflectance points within the $1.55~\mu$m band. We then used 

\begin{equation}
D_W = 100(1-(R_B/R_C)) 
\end{equation}

to determine the band depth as a percentage relative to the modeled linear continuum. Uncertainties were propagated using standard methods \citep[see][]{2010Uncertainties..H} such that

\begin{equation}
{\delta}D_W=100\left(\frac{R_B}{R_C}\right)\left(\left(\frac{{\delta}R_B}{R_B}\right)^2+\left(\frac{{\delta}R_C}{R_C}\right)^2\right)^\frac{1}{2}. 
\end{equation}

Ideally, we would have used the region of the spectrum at $1.25-1.35~\mu$m to define the coninuum on the short wavelength side of the absorption band, but the presence of an instrument artifact at $1.25-1.3~\mu$m prevented us from doing so (see Appendix). Defining the continuum using the spectrum at $1.15-1.25~\mu$m, a region with lower reflectance than continuum regions closer to the band, resulted in our estimate of $D_W$ being a lower limit. Ultimately we estimate that the band depth is $D_W=9.3\pm4.9\%$, which corresponds to a detection significance of $1.9\sigma$. The S/N of the spectrum was insufficient for us to search for the $1.65~\mu$m crystalline water ice band.

We attempted to predict the depth of the $2.0~\mu$m band we would have measured in our X-Shooter spectrum of 2012~DR$_{30}$ and compare it to the $12\pm1\%$ depth measured for a similar band observed by \citet{2018AJ....155..170S}. To do this we tried to predict the $2.0~\mu$m band depth based on the measured depth of the $1.55~\mu$m band and estimates of the band depth ratio between the $2.0~\mu$m and $1.55~\mu$m bands observed in the model reflectance spectra of 29 TNOs and centaurs reported by \citet{2008AJ....135...55B} and \citet{2012AJ....143..146B}; the selected objects had $H>4$ and an unambiguous detection of water ice. We followed the methods described in these publications to model the NIR spectra of our sample as the linear combination of a sloped continuum and a water ice spectrum such that

\begin{equation}
s(\lambda) = {f_W}{s_W}+(1-f_W)(m_C(\lambda-1.74)+0.49),
\end{equation} 

where $m_C$ is the gradient of the continuum, $s_W$ is a model water ice spectrum, and $f_W$ describes the proportional contribution of the water ice spectrum to the final model, $s(\lambda)$. To calculate $s(\lambda)$ we used $m_C$ and $f_W$ values reported by \citet{2012AJ....143..146B}, and used \citet{hapke_2012} theory to generate model water ice spectra, $s_W$. When modeling our water ice spectra we adopted the same functions used by \citet{2017Icar..287..218P}, although for the sake of simplicity we neglected to account for large scale surface roughness ($S(i,e,g)$). We adopt values of $B=1$, $h=0.025~$radians, and $\xi=0.3$ for the amplitude of the opposition effect, the angular width of the opposition effect, and the cosine asymmetry factor, respectively. $B$, $h$, and $\xi$ have never been measured for 2012~DR$_{30}$, so we selected values analogous to those measured for dynamically excited TNOs by the NASA \textit{New Horizons} probe \citep{2016ApJ...828L..15P,2019AJ....158..123V}. All our model reflectance spectra are compared to that of 2012~DR$_{30}$ in Figure \ref{fig:mod}.

We measured the average $2.0~\mu$m to $1.55~\mu$m water ice band depth ratio for the models generated while varying three parameters, including temperature¸ average grain size, and phase angle. In all tests we used water ice absorption coefficients measured in the lab by \citet{1998JGR...10325809G}, but swapped between those measured at 50~K, the temperature originally assumed for the models of \citet{2008AJ....135...55B} and \citet{2012AJ....143..146B}, and 90~K, which is close to the 96.3~K and 99.9~K subsolar temperatures of 2012~DR$_{30}$ when it was respectively observed by us with X-Shooter and by \citet{2018AJ....155..170S}. Because the rotation rate of 2012~DR$_{30}$ is unconstrained \citep{2013AA...555A...3K}, we consider these temperatures to be upper limits, hence our choice of absorption coefficients measured at 90~K rather than 100~K. We varied average grain sizes between $50~\mu$m, $16~\mu$m, and $5~\mu$m, inlcuding the original value assumed by \citet{2008AJ....135...55B} and \citet{2012AJ....143..146B}, and the best-fit values determined for water ice in the spectral models reported for 2012~DR$_{30}$ by \citet{2018AJ....155..170S}. Phase angles were swapped between $2.4^{\circ}$ and $3.6^{\circ}$, the phase angles of the NIR observations reported respectively by \citet{2018AJ....155..170S} and this work.

The depths of the $1.55~\mu$m and $2.0~\mu$m bands in our model water ice spectra were measured using the same method described previously for the $1.55~\mu$m band in our X-Shooter spectrum. For the $1.55~\mu$m bands in the model spectra, continuum regions were defined at $1.15-1.25~\mu$m and $1.7-1.8~\mu$m and a linear continuum was drawn between them. The average reflectance of each model spectrum was determined in the range $1.5-1.6~\mu$m, and that reflectance was compared to the reflectance of the linear continuum in order to measure the band depth as a percentage. Likewise continuum regions of $1.7-1.8~\mu$m and $2.22-2.26~\mu$m were used as continuum regions adjacent to the $2.0~\mu$m band, and the reflectance at $2.0-2.06~\mu$m was averaged to measure its depth.

\begin{table}
\centering
\caption{Modeled $2.0~\mu$m to $1.55~\mu$m Band Depth Ratios}
\begin{tabular}{c|ccc}
\hline\hline
~ & $5~\mu$m & $16~\mu$m & $50~\mu$m \\
\hline
50~K & 1.04|1.06 & 1.16|1.17 & 1.26|1.26\\
\hline
90~K & 1.04|1.04 & 1.17|1.17 & 1.28|1.28\\
\hline
\end{tabular}\\[6pt]
\small{\textbf{Note.} Table presenting the ratios of the $2.0~\mu$m band depth to the $1.55~\mu$m band depth averaged across 29 model reflectance spectra generated as described in section \ref{sec:nir} considering a variety of temperatures, grain sizes, and phase angles. In this table different rows record results for different temperatures, columns record results for different average grain sizes, and each data entry has two values corresponding to a result determined at a phase angle of $2.4^{\circ}$ on the left and at $3.6^{\circ}$ on the right. All values have standard errors of the mean at $\sim0.02$. Only changes in grain size cause a meaningful change to the band depth ratio across the parameter space tested.}
\label{tab:modres}
\end{table}

Measured average band depth ratios are presened in Table \ref{tab:modres}. We find that the average ratio of the $2.0~\mu$m band depth to that of the $1.55~\mu$m band in our sample of model reflectance spectra is invariant to temperature and phase angle within the uncertainties at the temperatures and angles that were tested. Changing the average grain size of the models had the greatest effect, with the band depth ratio increasing from ~1.05 to ~1.27 as grain size was decreased from  $50~\mu$m to $5~\mu$m. With these band depth ratios and the measured $\sim9\%$ depth of the $1.55~\mu$m band in the X-Shooter spectrum, we predict that, had we been able to extract it, the associated $2.0~\mu$m water ice band would have had a depth of between $9.5\%$ and $11.4\%$. This range overlaps with the $12\pm1\%$ $2.0~\mu$m band depth measurement of \citet{2018AJ....155..170S} within its uncertainties.

\subsection{The NUV Reflectance Drop}
We detect a deviation from linearity in our X-Shooter and FORS2 spectra of 2012~DR$_{30}$ that manifests as a drop in reflectance that curves downward from $\lambda\sim0.6~\mu$m towards NUV wavelengths. It is noticable in the greater spectral gradients measured at shorter optical wavelengths compared to longer ones; the gradient measured in the range $0.440-0.551~\mu$m is almost double that measured at $0.551-0.803~\mu$m (see Table \ref{tab:grad}). To quantify the significance of the deviation of the spectrum from a linear slope we performed a linear fit to the optical continuum of the spectrum in the range $0.6-0.9~\mu$m and extrapolated it to shorter wavelengths as shown in Figure \ref{fig:vis}. We then compared the average reflectance of 2012~DR$_{30}$ within the range $0.375-0.425~\mu$m in each of our spectra to that of the linear slope at $0.4~\mu$m; we take the standard deviation of the reflectance points in the same range as the uncertainty of the average reflectance at $0.4~\mu$m. From our X-Shooter spectrum we find that 2012~DR$_{30}$ is $13\pm4\%$ less reflective at $0.4~\mu$m than it would be if its spectrum had constant gradient across the full optical range, and consequently that the spectrum's deviation from linearity at this wavelength is detected at a significance of $3.3\sigma$. Identical analysis of our FORS2 spectrum shows that it is $16\pm3\%$ less reflective than the linear profile at $0.4~\mu$m, and that the spectrum's deviation from linearity at this wavelength is detected at $4.9\sigma$ significance. The measurements of the deviation of the spectra from linearity are consistent within their uncertainties.  

\section{Results}\label{sec:res}
In this section we discuss our results. First, we compare our reflectance spectra of 2012~DR$_{30}$ to each other and the published datasets presented in section \ref{sec:analysis}. Second, we compare the reflectance properties of 2012~DR$_{30}$ to other populations of minor planet. Third, we attempt to use our new data to constrain the surface composition of 2012~DR$_{30}$.

\subsection{Surface Variability}\label{sec:var}
We find that our two reflectance spectra of 2012~DR$_{30}$ are fully consistent with one another within their shared wavelength coverage, and that the spectral gradients measured from each are also consistent within their uncertainties (see table \ref{tab:grad}).

Without the uncertainties of the optical reflectance spectrum published by \citet{2018AJ....155..170S} it was not possible for us to determine $\sigma$ for our measurement of its gradient; this precluded robust quantitative comparison of that gradient to those measured from other datasets. Qualitative comparisons suggest that the shapes of our spectra and theirs are consistent in a broad sense, given that theirs also appears slightly redder at short optical wavelengths than it does at longer ones (see Table \ref{tab:grad}). We recognise, however, that the optical spectrum from \citet{2018AJ....155..170S} and all but one of the photometric datasets \citep[that reported by][]{2021AA...647A..71H} are formally inconsistent with our own spectra at short wavelengths, showing hardly any increased reddening at $\lambda<0.6~\mu$m, if any at all. That said, the photometric data do not necessarily have the precision to reveal such a feature, and moreover, are inconsistent between datasets. The spectrum from \citet{2018AJ....155..170S} may suffer from calibration errors of the order a few percent that would be sufficient to mask such a feature. In particular we note that the Solar calibrator they used has a precisely measured $U-B$ color reported in the SIMBAD database \citep[$U-B=0.255\pm0.002$;][]{2000A&AS..143....9W} that is redder than the $U-B$ color of the Sun \citep[$(U-B)_\odot=0.158\pm0.009$;][]{2012ApJ...752....5R} which suggests that Solar metal lines in their spectrum of 2012~DR$_{30}$ may have been overcorrected, producing a reflectance spectrum with decreased redness at short wavelengths. Given the remarkable consistency between the FORS2 and X-Shooter spectra, which were acquired at different epochs separated by almost two years and used different Solar calibrators \citep[one of which is a well characterised Solar twin;][]{2015A&A...574A.124D}, we trust that our spectra capture the true reflectance behaviour of 2012~DR$_{30}$. Self-consistent rotationally resolved observations of 2012~DR$_{30}$ at NUV/blue wavelengths will be necessary to determine whether the variability in the reported photometric data is real. 

Our comparison of the red optical reflectance data for 2012~DR$_{30}$ strongly suggests that its color variation is smaller than previously reported, or possibly entirely absent. When they are measured with consistent methods at $0.55-0.8~\mu$m and compared, we find that the spectral gradients of four of the seven datasets presented in Figure \ref{fig:vis} and table \ref{tab:grad} fall within the range $10\pm1~\%/0.1~\mu$m, and the gradient measured from the SDSS data is also consistent with $\sim10~\%/0.1~\mu$m within its uncertainties; the only inconsistent reflectance data is that derived from the CQUEAN photometry published by \citet{2018AJ....155..170S}, being significantly redder than the rest. Detections of red spots or patches on the surfaces of icy minor planets are not unprecedented \citep[e.g.][]{2009AJ....137.3404L,2015ApJ...804...31F,2018NatAs...2..133F} and it is possible that \citet{2018AJ....155..170S} detected something similar, but such a scenario seems unlikely. The greater weight of evidence suggesting that 2012~DR$_{30}$ has a fairly constant surface color and its previously reported lack of rotational lightcurve \citep{2013AA...555A...3K} instead indicate that the surface of 2012~DR$_{30}$ is likely to be much more homogenous than was previously suggested.

Despite the fact that our NIR spectrum of 2012~DR30 and that reported by \citet{2018AJ....155..170S} test positive for signatures of water ice, they are more different than alike. The FIRE spectrum is significantly redder than the NIR X-Shooter spectrum, and also lacks the $1.55~\mu$m water ice absorption that we observed. The presence of a $2.0~\mu$m water ice absorption band cannot be confirmed in the X-Shooter spectrum, but we predict its presence based on our tentative detection of the weaker $1.55~\mu$m band. There are a number of unknown quantities in our spectral modeling analysis (see section \ref{sec:nir}), particularly the average size of the surface grains on 2012~DR$_{30}$, so we only use the results of that analysis to give a general impression of the variablility of the water ice band strength. It appears likely that the $9.5-11.4\%$ deep $2.0~\mu$m band predicted to be in the X-Shooter spectrum is no stronger than the $12\pm1\%$ deep band reported by \citet{2018AJ....155..170S}. If the band depth is closer to the lower end of the predicted range, the variation of the band strength between observing epochs is consistent with that previously reported for the reflectance spectra of TNOs and centaurs \citep[e.g.][]{2005A&A...444..977M,2009A&A...501..375P,2010AJ....140.2095B,2015ApJ...804...31F}, and therefore seems unremarkable. Given the non-dectection of a lightcurve for 2012~DR$_{30}$ reported by \citet{2013AA...555A...3K} and the following implication that its albedo does not appear to vary, we lean towards considering it more likely that the abundance of water ice on 2012~DR$_{30}$ is fairly constant across its surface.  

Within the uncertainties of their measurement, and the range of band depth ratios estimated by our spectral modeling work, we'd expect that the $2.0~\mu$m water ice band detected by \citet{2018AJ....155..170S} would be accompanied by a $1.55~\mu$m band with a depth of 8.7-12.4\%, but they did not observe one. The variable presence of the $1.55~\mu$m water ice band is curious, especially if the surface abundance of water ice is not variable. If this is the case, and if the $1.55~\mu$m band variation is real, it may indicate a variegation of the surface's physical properties such as grain size distribution or porosity. Like \citet{2018AJ....155..170S}, we note that there have been reports of TNOs with spectra that have only the $2.0~\mu$m water ice band and lack the $1.55~\mu$m band \citep{2014A&A...562A..85L}. With only two epochs of the NIR data available for 2012~DR$_{30}$, however, the extent to which its NIR spectrum varies is presently unconstrained; more self consistent NIR observations are needed before a better understanding of the properties of the surface water ice on 2012~DR$_{30}$ can be obtained.     

\subsection{Comparison to Other Minor Planets}\label{sec:comp}
\begin{figure}
\centering
\includegraphics[scale=0.7]{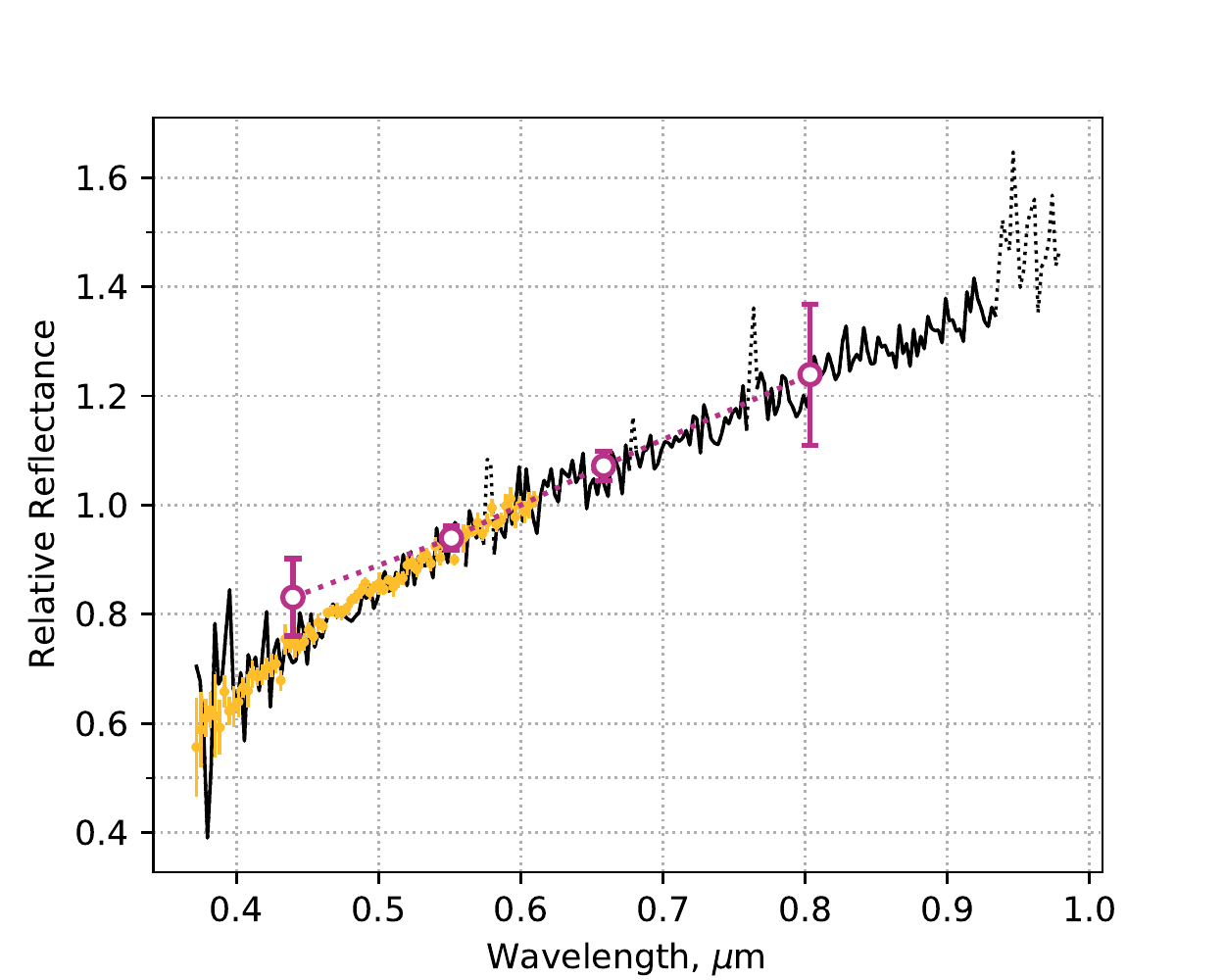}
\caption{This figure compares our optical reflectance data for 2012~DR$_{30}$ to the average reflectance properties of high inclination objects. The X-Shooter and FORS2 reflectance spectra are respectively plotted in black and yellow, while a coarse reflectance spectrum derived from the average $BVRI$ colors of 21 high inclination objects reported by \citet{2021AA...647A..71H} is plotted in red. The average colors used to create this plot are $B-V=0.79\pm0.07$, $V-R=0.50\pm0.05$, and $R-I=0.50\pm0.12$. \label{fig:ave}}
\end{figure}

We find that 2012~DR$_{30}$ is moderately red in color with an optical spectral gradient of $10\pm1~\%/0.1~\mu$m at $0.55-0.8~\mu$m, which makes the color of 2012~DR$_{30}$ consistent with the color distributions of a number of other populations such as the Damocloids, the Trojan asteroids of Jupiter and Neptune, the irregular satellites of the giant planets, the grey centaurs, the D-type asteroids, the comets, and the less-red TNOs \citep[e.g.][]{2010AJ....139.1394S,2015AJ....150..201J,2018A&A...618A.170L}. The water ice absorption bands and grey NIR continuum of our X-Shooter spectrum of 2012~DR$_{30}$ are observed fairly frequently in the NIR spectra of centaurs and TNOs of a similar size \citep[e.g.][]{2009Icar..201..272G,2011Icar..214..297B,2012AJ....143..146B}.

By taking the average $BVRI$ colors of all 21 high inclination objects reported in Table 4 of \citet{2021AA...647A..71H} and plotting the resulting coarse reflectance spectrum over our spectra of 2012~DR$_{30}$ we find that the color of 2012~DR$_{30}$ is completely consistent with the average color of high inclination objects at $0.55-0.8~\mu$m (see Figure \ref{fig:ave} for details); the gradient of the coarse average spectrum in this range is $11\pm4~\%/0.1~\mu$m. At $0.44-0.55~\mu$m, however, 2012~DR$_{30}$ appears redder than the average spectrum with respective spectral gradients of $19.7\pm0.9~\%/0.1~\mu$m and $11\pm8~\%/0.1~\mu$m, but note that within their uncertainties these values are still borderline consistent. Therefore, while linearity appears to be the norm for the optical reflectance spectra of most high inclination Solar System objects \citep[e.g.]{2013A&A...550A..13P,2018A&A...618A.170L,2021AA...647A..71H}, there appears to be sufficient spread in their average $B-V$ color that some may indeed have NUV reflectance drops that have not yet been confirmed (see Figure \ref{fig:ave}). As a result we can neither rule out the possibility that 2012~DR$_{30}$ is an oddball with an unusually red color at $\lambda<0.55~\mu$m, nor that it is one of a subclass of high inclination objects with reddened NUV reflectance spectra.    

\begin{table*}
\centering
\caption{\textit{HiHq} Centaur Spectral Gradients in $\%/0.1~\mu$m}
\begin{tabular}{p{0.6cm}p{0.5cm}p{0.5cm}p{1.7cm}p{3.cm}p{1.cm}p{1.75cm}}
\hline\hline
$S'$ & $\sigma$ & $\sigma_{S'}$ & $\lambda$~Range\textsuperscript{a} & Object & Bands\textsuperscript{c} & Reference\textsuperscript{d} \\[2pt]
\hline
4.1   & 2.1 & 0.1 & 0.551-0.803 & 127546 (2002~XU$_{93}$) & VRI & (1,2)\\[1pt]
6.6   & 3.3 & 0.1 & 0.551-0.803 & 528219 (2008~KV$_{42}$) & VRI & (1)\\[1pt]
14.2  & 2.4 & 0.1 & 0.551-0.803 &         2010~WG$_{9}$   & VRI & (3)\\[1pt]
7.2   & 2.0 & 0.1 & 0.551-0.803 &         2010~WG$_{9}$   & VRI & (3)\\[1pt]
1.2   & 2.4 & 0.1 & 0.551-0.803 &         2010~WG$_{9}$   & VR  & (4)\\[1pt]
\hline
8.8   & 3.7  & 0.1 & 0.440-0.551 & 127546 (2002~XU$_{93}$) & BV & (2)\\[1pt]
14.3  & 10.7 & 0.3 & 0.440-0.551 & 528219 (2008~KV$_{42}$) & BV & (1)\\[1pt]
12.0  & 3.4  & 0.1 & 0.440-0.551 &         2010~WG$_{9}$   & BV & (3)\\[1pt]
6.7   & 5.8  & 0.1 & 0.440-0.551 &         2010~WG$_{9}$   & BV & (4)\\[1pt]
\hline
\end{tabular}\\[6pt]
\small{\textbf{Note.} Here we present spectral gradients ($S'$), associated $1\sigma$ uncertainties ($\sigma$), and associated standard errors ($\sigma_{S'}$) measured for three \textit{HiHq} centaurs from their published photometric colors. Literature data taken from \citep[1;][]{2010AJ....139.1394S}, \citep[2;][]{2012AA...546A.115H}, \citep[3;][]{2013AJ....146...17R}, and \citep[4;][]{2015AJ....150..201J}. $VRI$ colors of 2010~WG$_{9}$ from \citet{2013AJ....146...17R} labelled VRI\textsuperscript{1} and VRI\textsuperscript{2} were observed at different rotational phases.}\\
\label{tab:gradhihq}
\end{table*}

Similar to the $BVRI$ colors of 2012~DR$_{30}$ presented in Table \ref{tab:grad}, those published for three $HiHq$ centaurs were converted to reflectance and used to measure their optical spectral gradients, which are presented in Table \ref{tab:gradhihq}. As previously reported, and as is typical of high inclination objects \citep[e.g.][]{2021AA...647A..71H}, the colors of 2002~XU$_{93}$, 2008~KV$_{42}$, and 2010~WG$_{9}$ are generally grey to moderately red \citep{2010AJ....139.1394S,2012AA...546A.115H,2013AJ....146...17R,2015AJ....150..201J}. It is interesting to note that each $HiHq$ centaur in table \ref{tab:gradhihq} has at least one set of colors where the gradient measured at $\lambda<0.55~\mu$m is at least double that measured at longer wavelengths, suggesting the presence of a reflectance drop at short wavelengths similar to that observed in our reflectance spectra of 2012~DR$_{30}$. The caveat to this observation, however, is that within their uncertainties the gradients measured at short and long wavelengths are consistent, which means that we cannot rule out the possibility that the optical reflectance spectra of 2002~XU$_{93}$, 2008~KV$_{42}$, and 2010~WG$_{9}$ are linear across the full optical range. If all \textit{HiHq} centaurs do in fact become redder from optical to NUV wavelengths, the low frequency of such features in the spectra of low inclination centaurs and TNOs would support the inference made from dynamical studies that the \textit{HiHq} centaurs are not sourced from the TNO population \citep[see][]{2012MNRAS.420.3396B,2013Icar..224...66V}. A shared general spectral shape for the \textit{HiHq} centaurs suggests that they have common surface properties, and that they may have either formed under similar conditions or experienced similar types and extents of surface processing. Alternatively a NUV reflectance drop may be acquired only by objects which have very large aphelion distances that cause them to spend long periods of their orbit beyond the heliosheath; at these great heliocentric distances organics on their surfaces would likely be heavily dehydrogenated and aromatised as a result of strong irradiation by galactic cosmic rays \citep[e.g.][]{2008ssbn.book..507H}. Precise spectroscopic observations of more \textit{HiHq} centaurs at NUV and optical wavelengths may be able to confirm the presence of a NUV reddening in their spectra.

Differences in spectral gradient between the NUV and optical regions are not commonly observed in the reflectance spectra of objects like centaurs and TNOs \citep[e.g.][]{2004A&A...421..353F,2009A&A...508..457F}. Such a feature has been clearly identified in the spectrum of the TNO 120216~(2004~EW$_{95}$), but this unusual object is predicted to be a carbonaceous (C-type) asteroid interloping in the trans-Neptunian belt \citep{2018ApJ...855L..26S}. Beside this, only the centaur 32532~Thereus has a reported spectrum that hints at the presence of a NUV reflectance drop \citep{2002A&A...392..335B}; the feature is only detected tentatively, however, and has a different shape to the reflectance drop that 2012~DR$_{30}$ has. With an orbital inclination $i=20.4^{\circ}$, Thereus is not a $HiHq$ centaur, and unlike 2012~DR$_{30}$ it probably originated among the TNOs.  NUV reddening is more common in the spectra of main belt asteroids \citep[see][]{2009Icar..202..160D}, and is also observed in the reflectance spectra of the less-red Jupiter Trojan asteroids \citep[at $\lambda<0.5~\mu$m;][]{2019AJ....157..161W}. The presence of this feature in the reflectance spectrum of 2012~DR$_{30}$ therefore tentatively hints that this object may have more in common with those popultations than it does with the TNOs.

\subsection{Surface Composition}
The only clear assertion we can make about the surface composition of 2012~DR$_{30}$ is that it has surface water ice. Both of the characteristic $1.55~\mu$m and $2.0~\mu$m water ice absorption bands have been observed in reflectance spectra of 2012~DR$_{30}$, although at different epochs \citep[see section \ref{sec:analysis} and ][]{2018AJ....155..170S}. Like \citet{2018AJ....155..170S} we have not detected the $\sim0.9~\mu$m absorption band predicted to be present by \citet{2013AA...555A...3K}. 

The surface material responsible for the NUV reflectance drop in the spectrum of 2012~DR$_{30}$ is not clearly identifiable, but a process of elimination can narrow down the list of candidates. Water ice, although detected in the NIR, does not absorb strongly at NUV/blue wavelengths and can be ruled out. For the same reason, and because its characteristic absorption bands at $2.27~\mu$m and $2.34~\mu$m have not been detected \citep[see][]{2018AJ....155..170S}, methanol ice can also be ruled out.

Anhydrous mafic silicates such as olivine and pyroxene have strong NUV absoprtion features in their reflectance spectra \citep[e.g.][]{2015aste.book...43R}, but if anhydrous silicates are present in sufficient quantity on the surface of 2012~DR$_{30}$ to be the cause of the observed reddening in its spectrum, we would expect their other strong chracteristic NIR absorption bands to also be detectable. These bands would appear at $\lambda\sim1.0~\mu$m for olivine, and at $\lambda\sim1.0~\mu$m and $2.0~\mu$m for pyroxene, but detections of such bands have never been confirmed for 2012~DR$_{30}$. The addition of amorphous carbon (AC) to a silicate surface has been shown to weaken these bands, but not without also weakening the NUV feature to a similar extent \citep{2011Icar..212..180C}, so the appearance of only the NUV reflectance drop appears inconsistent with a mixture of silicate and AC. It therefore seems reasonable to exclude anhydrous mafic silicates as contributors to the NUV reddening in the spectrum of 2012~DR$_{30}$. Notably, however, model reflectance spectra of mixtures of materials present in chondritic porous interplanetary dust particles (IDPs) have been shown to match well with those of the dark, moderately red, and icy asteroids to which 2012~DR$_{30}$ appears similar in color and albedo. Since some of the dominant phases in IDPs are olivine, enstatite, and amorphous silicates, we cannot completely rule out the presence of anhydrous silicates on its surface \citep[see][]{2015ApJ...806..204V}.   

The low albedo and moderately red optical color of 2012~DR$_{30}$ suggests that its surface may be at least partially comprised of macromolecular refractory organics of the kind produced through radiolytic and photolytic processing of simple molecular ices \citep[e.g.][]{2006ApJ...644..646B}. Refractory organics containing aromatic rings (such as Polycyclic Aromatic Hydrocarbons; PAHs) and other $\pi$-bonded molecular structures are well documented to absorb light strongly at NUV wavelengths and have strong NUV absorption features in their spectra that advance to longer wavelengths as the size and clustering of their $\pi$-bond networks increase \citep[e.g.][]{2014Icar..237..159I}. If the NUV reddening in the spectrum of 2012~DR$_{30}$ is produced by aromatic refractory organics on its surface, it is plausible that its infrared spectrum will exhibit the $3.3~\mu$m fundamental aromatic C-H stretching mode that is typically used to identify the presence of aromatic organics on Solar System objects \citep[e.g.][]{2014Icar..233..306C}. No spectroscopic observations of 2012~DR$_{30}$ have yet been reported in the $3-4~\mu$m range, but if future observations at these wavelengths do detect an aromatic feature at $3.3~\mu$m we can be more confident that the NUV reflectance drop is at least partly the result of absorption by aromatic hydrocarbons. The higher harmonics of the $3.3~\mu$m C-H stretch band are observed in laboratory samples of PAHs at $\sim1.687~\mu$m, $\sim1.145~\mu$m, and $\sim0.880~\mu$m \citep{2014Icar..237..159I}, but these are not observed in our spectrum of 2012~DR$_{30}$. This shouldn't be surprising, however, as these features are easily washed out if a specific PAH species is not observed in a highly pure form \citep{2014Icar..237..159I}. It is likely that any organics on the surface of 2012~DR$_{30}$ are a diverse mixture rather than a single uniform species, and they are also definitely mixed with other materials. For this reason we cannot claim that our non-detection of NIR C-H stretch modes is evidence against the presence of aromatic organics on the surface of 2012~DR$_{30}$.

Drops in reflectance observed in the spectra of dark carbonaceous main-belt asteroids are often attributed to absorption by Fe\textsuperscript{2+}$\,\to\,$Fe\textsuperscript{3+} intervalence charge transfer (IVCT) transitions occuring in iron oxide minerals contained within iron-bearing phyllosilicates on their surfaces \citep[e.g.][]{1994Icar..111..456V,2015aste.book...65R}; the detection of such a NUV reflectance drop is not diagnostic, however. Other optical phyllosilicate bands such as the weak $0.7~\mu$m Fe\textsuperscript{2+}$\,\to\,$Fe\textsuperscript{3+} IVCT band are easily obscured by carbonaceous material \citep[e.g.][]{2011Icar..212..180C}, so if the NUV reddening in the spectrum of 2012~DR$_{30}$ is caused by ferric products of aqueous alteration it is not surprising that their other optical bands have not been detected. It should also be noted that carbonaceous material is capable of reducing the strength of the NUV reflectance drops observed in the spectra of aqueously altered asteroids \citep{2019GeoRL..4614307H}, so if the reflectance drop is caused by aqueously altered minerals on the surface of 2012~DR$_{30}$ the additional presence of carbon rich material may explain why its NUV reddening is not as strong as that observed in the spectra of pure phyllosilicate samples or aqueously altered meteorites. Confirmation that the NUV reflectance drop is caused (even in part) by the presence of aqueously altered minerals on the surface of 2012~DR$_{30}$ may be achieved by the detection of a strong $3.0~\mu$m absorption band associated with structural hydroxyl (OH) or adsorbed water contained within them \citep[e.g.][]{2012Icar..219..641T,2015aste.book...65R}.

\section{Conclusions}
We have presented two new reflectance spectra of the extreme $HiHq$ centaur 2012~DR$_{30}$, which combined have a wavelength coverage of $0.37-1.80~\mu$m; through analysis of these spectra we have attempted to constrain the color and surface composition of 2012~DR$_{30}$. 

Contrary to previous reports of the optical color variation of 2012~DR$_{30}$, we found through consistent measurements of our spectra and several previously published spectra and colors that its spectral gradient at $0.55-0.8~\mu$m appears consistent with $S'\simeq10\pm1~\%/0.1~\mu$m across almost all observational epochs reported so far. The surface homogeneity that this finding suggests is consistent with the report by \citet{2013AA...555A...3K} that 2012~DR$_{30}$ exhibits a minimal rotational lightcurve and therefore also has minimal variation in albedo across its surface. Two epochs of published photometry, if taken at face value, hint that 2012~DR$_{30}$ may have red spots on its surface (see section \ref{sec:var}), but self-consistent rotationally resolved color measurements will be required to confirm this.

The homogeneity of the reported optical colors of 2012~DR$_{30}$ does not seem to extend to NUV/blue wavelengths, however. There is now considerable variation in the reported colors of 2012~DR$_{30}$ at $\lambda<0.55~\mu$m, which is at odds with the finding that the spectral gradients measured from our own spectra in this region are completely consistent. Spectral variation also appears at NIR wavelengths. We have tentatively confirmed the detection of water ice on 2012~DR$_{30}$ reported by \citet{2018AJ....155..170S}, but by a weak detection of a $1.55~\mu$m water ice band that was absent from their spectrum. Our spectrum does not extend to wavelengths above $1.8~\mu$m, so unfortunately we cannot confirm their detection of the $2.0~\mu$m water ice band. Based on the depth of the $1.55~\mu$m band and model water ice reflectance spectra we predict that the $2.0~\mu$m band strength has not changed much between the observations of \citet{2018AJ....155..170S} and ours. If the abundance of water ice is fairly constant across the surface of 2012~DR$_{30}$, the inconsistent appearance of the $1.55~\mu$m water ice band may hint that the physical properties of the surface of 2012~DR$_{30}$ are more strongly variegated than its composition. In both the NUV/blue and NIR cases, any observed variation in color may be at odds with the previously reported lack of rotational lightcurve for 2012~DR$_{30}$ \citep{2013AA...555A...3K}. Further self-consistent rotationally resolved observations of 2012~DR$_{30}$ at NUV/blue and NIR wavelengths are needed to determine if the observed variation is real, and if so what may be causing it.

We have consistently detected a NUV reddening in the reflectance spectrum of 2012~DR$_{30}$ at $\lambda<0.6~\mu$m. Such features are rare in the reflectance spectra of TNOs, supporting arguments from dynamical studies that suggest 2012~DR$_{30}$ and other $HiHq$ centaurs are not derived from the trans-Neptunian scattered disk or the Kuiper belt. That said, strong evidence of other high inclination minor planets with a similarly shaped optical reflectance spectrum is currently lacking, although future spectroscopic observations targeting high inclination objects may change that. The cause of the NUV reflectance drop is currently unclear, but we are able to rule out materials such as water ice, methanol ice, and anhydrous mafic silicates as likely candidates. Instead we favour the possibility that the reflectance drop may be caused by aromatic hydrocarbons and/or ferric oxides contained within aqueously altered silicates. Confirmation of the presence of these materials, however, will require observations of 2012~DR$_{30}$ at wavelengths of $3-4~\mu$m.

\acknowledgments

We are grateful to Emanuela Pompei, Evelyn Johnston, and Stephane Brillant for sharing their observing expertise and assisting our observations at the VLT. Thanks also to Faith Vilas for sharing her copy of Brian Skiff's list of solar calibrator stars. T.S. is supported through a Gemini Science Fellowship by the international Gemini Observatory, a program of NSF’s OIR Lab, which is managed by the Association of Universities for Research in Astronomy (AURA) under a cooperative agreement with the National Science Foundation, on behalf of the Gemini partnership of Argentina, Brazil, Canada, Chile, the Republic of Korea, and the United States of America. T.S. was also supported in part by the Astrophysics Research Centre at Queen's University Belfast, and the Northern Ireland Dept.~for the Economy. This work is based on observations collected at the European Organisation for Astronomical Research in the Southern Hemisphere under ESO programs 093.C-0259, 095.C-0521 and 099.C-0651. This research made use of NASA’s Astrophysics Data System Bibliographic Services, the JPL HORIZONS web interface (https://ssd.jpl.nasa.gov/horizons.cgi), and data and services provided by the IAU Minor Planet Center.

%

\vspace{5mm}
\facility{ESO: VLT-UT1(FORS2), VLT-UT2(X-Shooter)}


\software{Astropy \citep{2013A&A...558A..33A}, ESO Reflex \citep{2013A&A...559A..96F}, Matplotlib \citep{2007CSE..9..90H}, NumPy \citep{harris2020array}, SciPy \citep{2020SciPy-NMeth}}



\newpage

\begin{figure}
\centering
\includegraphics[scale=0.56]{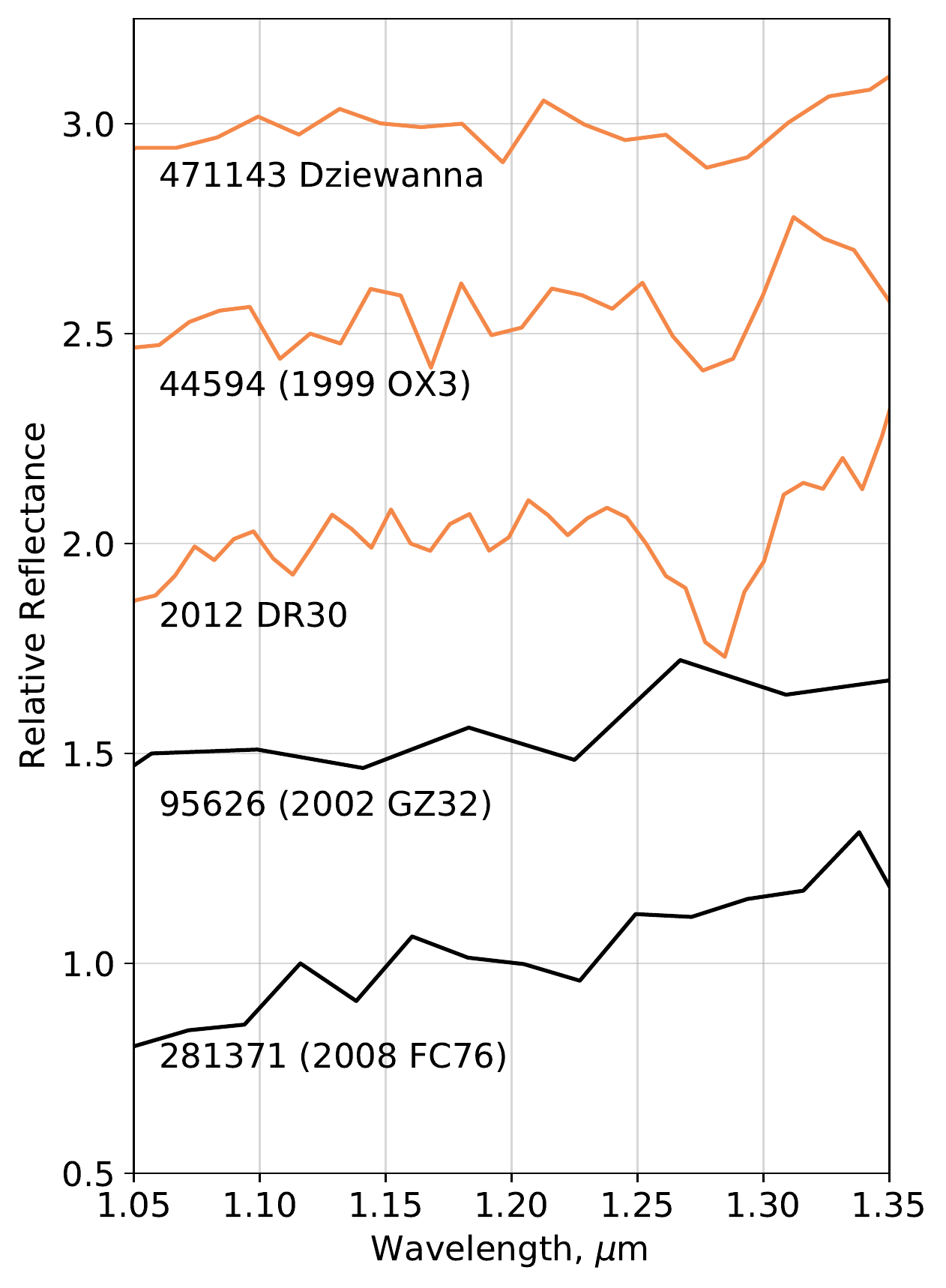}
\caption{A sample of NIR reflectance spectra of TNOs and centaurs observed by our team with X-Shooter. In black we plot reflectance spectra observed as part of ESO program 093.C-0259 and in red we plot those observed as part of ESO program 095.C-0521, which respectively did and did not make use of X-Shooter's $K$-band blocking filter. All these spectra have been scaled to unity at $\lambda=1.15~\mu$m and are offset in increments of +0.5 for clarity.  \label{fig:artifact}}
\end{figure}

\appendix
Here we discuss an instrumental artifact in our NIR X-Shooter spectrum of 2012~DR$_{30}$, which manifests as a strong and fairly sharp feature at $1.25-1.30~\mu$m. Comparison of NIR spectra of faint TNOs and centaurs that we have observed with X-Shooter in Figure \ref{fig:artifact} demonstrates that the artifact appears only in spectra observed without use of the $K$-band blocking filter. As all the presented spectra have been calibrated with a different Solar calibrator star, we also know that the choice of star doesn't have any bearing on whether the artifact appears or not. ESO's X-Shooter instrument manual documents how observations that do not use the $K$-band blocking filter have significantly increased thermal background noise in the $J$- and $H$-bands caused by scattered thermal radiation from $K$-band. In addition, the S/N of the spectra are further reduced in the affected region at $1.25-1.30~\mu$m by simultaneously containing multiple moderately strong sky emission lines and falling at the join between NIR echelle orders 20 and 21, rather than at the center of an order where S/N would be higher. The artifact therefore appears to be the result of multiple factors combining to effectively negate the signal of the faint 2012~DR$_{30}$ at $1.25-1.3~\mu$m, while the strong signal of the much brighter solar calibrator was sufficient to overcome the higher background. The weaker spectrum from 2012~DR$_{30}$ was then overcorrected by the stronger  Solar calibrator spectrum at these wavelengths, producing a strong dip in the final reflectance spectrum. As the $1.25-1.3~\mu$m feature in the spectrum of 2012~DR$_{30}$ is not intrinsic to 2012~DR$_{30}$ itself it was not considered for analysis. We briefly note that both brighter targets and longer exposures are less affected by this artifact. For example the pre-outburst spectrum of 174P/Echeclus reported by \citet{2019AJ....157...88S} was observed without the $K-$band blocking filter but shows no sign of the $1.25-1.3~\mu$m artifact, likely because 174P was $\sim2$ magnitudes brighter than 2012~DR$_{30}$ at the time it was observed. An X-Shooter spectrum of a faint TNO, 278361 (2007~JJ$_{43}$), that was observed without the $K-$band blocking filter and reported by \citet{2015A&A...582A..13G} has a small $1.25-1.3~\mu$m artifact that appears to have been somewhat mitigated by the extremely long (1800~s) exposure times used by that team.

\newpage

\bibliography{masterbib}{}
\bibliographystyle{aasjournal}



\end{document}